\documentclass[sigconf]{acmart}

\usepackage{graphicx}
\usepackage{booktabs}
\usepackage{multirow}
\usepackage{multicol}
\usepackage{adjustbox}
\usepackage{url}
\usepackage{balance}
\usepackage{amsmath}
\usepackage{xspace}
\usepackage{enumitem}
\usepackage{csquotes}
\usepackage{tikz}

\AtBeginDocument{%
  }

\setcopyright{acmlicensed}
\copyrightyear{2025}
\acmYear{2025}
\acmDOI{
}
\acmConference[
] {
}
\acmBooktitle{
}
\acmISBN{
979-8-4007-0436-9/24/10
}
\begin{document}

\title{System-driven Interactive Design Support for Cloud Architecture: A Qualitative User Experience Study with Novice Engineers}

\author{Ryosuke Kohita\footnotemark[1]}
\affiliation{%
  \institution{CyberAgent}
  \city{Tokyo}
  \country{Japan}
}
\email{kohita_ryosuke@cyberagent.co.jp}

\author{Akira Kasuga\footnotemark[1]}
\orcid{0009-0003-8870-8010}
\affiliation{%
  \institution{CyberAgent, Inc.}
  \city{Tokyo}
  \country{Japan}
}
\email{kasuga_akira@cyberagent.co.jp}

\renewcommand{\shortauthors}{Ryosuke Kohita \& AKira Kasuga}

\def\eg{{\it e.g.}}
\def\cf{{\it c.f.}}
\def\ie{{\it i.e.}}
\def\etal{{\it et al. }}
\def\etc{{\it etc}}


\keywords{cloud architecture, interactive design support, system-driven approach, novice engineers, case study, educational technology}

\begin{abstract}
Cloud architecture design presents significant challenges due to the necessity of clarifying ambiguous requirements and systematically addressing complex trade-offs, especially for novice engineers with limited cloud experience. While recent advances in the use of AI tools have broadened available options, system-driven approaches that offer explicit guidance and step-by-step information management may be especially effective in supporting novices during the design process.
This study qualitatively examines the experiences of 60 novice engineers using such a system-driven cloud design support tool. The findings indicate that structured and proactive system guidance helps novices engage more effectively in architectural design, especially when addressing tasks where knowledge and experience gaps are most critical. For example, participants found it easier to create initial architectures and did not need to craft prompts themselves.
In addition, participants reported that the ability to simulate and compare multiple architecture options enabled them to deepen their understanding of cloud design principles and trade-offs, demonstrating the educational value of system-driven support.
The study also identifies areas for improvement, including more adaptive information delivery tailored to user expertise, mechanisms for validating system outputs, and better integration with implementation workflows such as infrastructure-as-code generation and deployment guidance. Addressing these aspects can further enhance the educational and practical value of system-driven support tools for cloud architecture design.
\footnote{Both authors contributed equally to this research.}
\end{abstract}

\maketitle

\section{Introduction}
\label{sec:intro}
Designing robust cloud architectures requires in-depth technical knowledge of cloud services as well as architectural expertise to integrate them into coherent systems.
This challenge becomes even greater when requirements are ambiguous or evolving~\cite{claus-2018-architectural}.
Novice engineers in particular often struggle to manage the broad spectrum of design considerations, systematically refine requirements, and evaluate alternative solutions~\cite{kumar2023modeling}.
Although chat-based AI tools offer flexible assistance, they place the burden of steering interactions on users~\cite{pereira2023}. As a result, important factors may be missed, leading to persistent uncertainty throughout the process~\cite{zi2025understand, diazpace2024helping}.

A promising alternative is the \emph{system-driven approach}, where explicit workflows guide the design process instead of leaving it open-ended. Such tools provide step-by-step assistance, helping users systematically navigate decisions and bridge knowledge gaps.
Prior work introduced a system-driven tool for cloud architecture design (CA-Buddy) \cite{kohita2025systemdrivencloudarchitecturedesign}. In a role-playing experiment with skilled engineers, participants rated the user experience of CA-Buddy highly, as it surfaced overlooked issues and reduced cognitive burden. These results suggest that system-driven approaches enhance thoroughness and user confidence, leading to more robust designs.

Recent work suggests significant promise in using system-driven approaches to help novice engineers overcome the unique challenges of cloud architecture design. Less experienced practitioners frequently lack the domain understanding necessary to clarify requirements, manage trade-offs, and anticipate pitfalls~\cite{diazpace2024helping}, making them especially prone to oversights and inefficiency.
By providing explicit guidance and bridging expertise gaps, system-driven tools could substantially improve outcomes for novices. However, empirical evidence in novice contexts remains limited, though explored in ML/DS education \cite{han2022mladder}

In this work, we present a qualitative study with 60 novice engineers using CA-Buddy to evaluate system-driven support for novices.
Structured and proactive system guidance was reported to reduce cognitive load in architecture design.
Many participants specifically noted benefits in areas where knowledge and experience gaps are most problematic, such as proposing initial architectures and avoiding the need to craft prompts themselves.
The educational value of the system was also emphasized; the ability to simulate and compare a variety of architectures enabled deeper learning of cloud design principles and trade-offs.
For practical adoption, participants pointed out the need for more adaptive information delivery, validation of outputs, and integration with deployment workflows.
Taken together, these results indicate that system-driven approaches not only improve conceptual design quality, but are also expected to further support learning and operational processes for less experienced engineers.

\begin{figure*}[!t]
  \centering
  \includegraphics[width=1.0\linewidth]{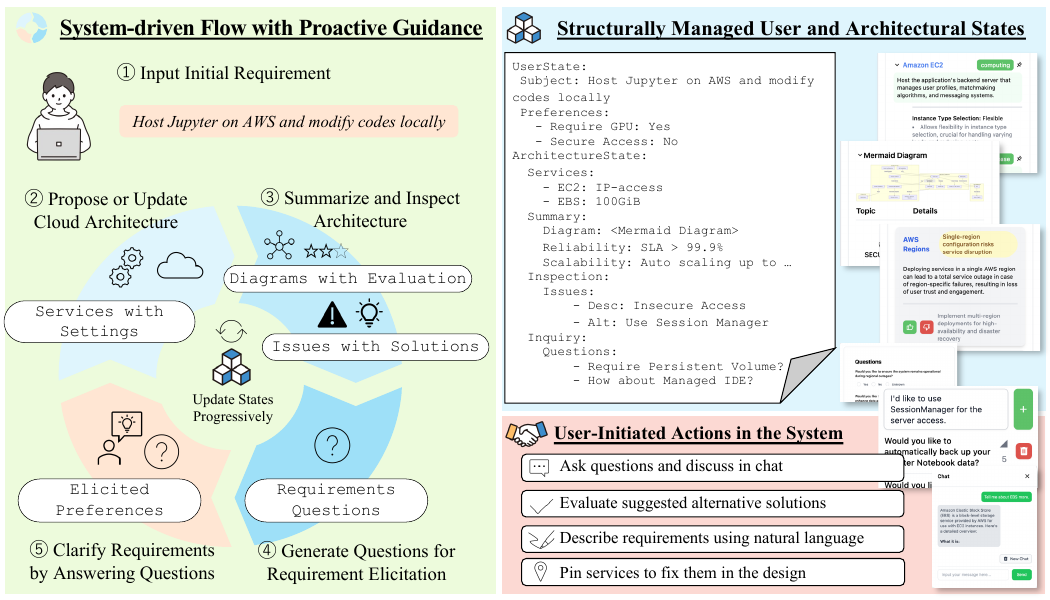}
  \caption{\small Overview of CA-Buddy, a workflow-based tool that guides users through cloud architecture design via a system-driven flow (left; green), structured state management (top right; blue), user actions (bottom right; red), and a UI snapshot (far right).}
  \label{fig:ca-buddy-overview}
\end{figure*}

\section{System Overview: CA-Buddy}
\label{sec:system}

In this study, we use CA-Buddy \cite{kohita2025systemdrivencloudarchitecturedesign}, a system-driven cloud architecture design support tool. CA-Buddy proactively guides users through a structured, workflow-based process, organizing requirements and design states to ensure consistency and traceability. This section outlines the system’s workflow-based guidance and its hybrid integration of chat-based interactions, as informed by prior research. Figure~\ref{fig:ca-buddy-overview}
 illustrates the overview.

\paragraph{System-driven Flow with Proactive Guidance}
CA-Buddy guides users through a systematic, iterative workflow for cloud architecture design. The process begins with users entering their initial requirements. Based on this input, the system generates a preliminary architectural proposal. CA-Buddy then summarizes the proposed architecture and automatically inspects it to identify any issues or concerns. When gaps or ambiguities remain, the system proactively poses targeted questions to the user, prompting further refinement of requirements and solutions. After receiving the user's responses, the workflow cycles back—enabling incremental improvements with each iteration. All system actions, including proposal generation, summarization, inspection, and inquiry, are powered by large language models (LLMs) that leverage task-specific prompts and contextual information to adapt to the user’s evolving needs.

\paragraph{Structured Management of User and Architecture States}
CA-Buddy maintains consistency and supports effective interaction by managing user requirements and architectural states in a structured manner. The system uses two core models: \textsf{UserState} and \textsf{ArchitectureState}.
\textsf{UserState} represents the main aspects of the project, such as use cases, goals, technical constraints, and user preferences. It also captures responses to system inquiries and feedback on alternative solutions, which guide subsequent system actions.
\textsf{ArchitectureState} tracks the current architecture configuration, including selected cloud services, a summary of the design, identified concerns or risks, and outstanding questions for users. These elements are continuously updated as the workflow progresses, ensuring that the support provided remains relevant to the user’s current context and requirements.

\paragraph{User-Initiated Actions}

Beyond the system-driven workflow, CA-Buddy allows users to directly influence the design process in several ways. Users can specify mandatory services to be included in the architecture, ensuring their preferences are reflected in future proposals. They can also provide feedback on suggested alternatives, marking options as favorable or unfavorable, which shapes subsequent recommendations. Additionally, users are able to ask questions or seek advice via a chat interface, and can give free-form design instructions using natural language. Notably, the chat-based consultation and natural language instructions were newly introduced in this study based on previous user feedback. While the workflow remains primarily system-driven, these capabilities offer greater flexibility for active exploration and user-driven iteration.

\section{User Experience Study with Novice Engineers}
This section presents a user study in which we collected qualitative feedback novice engineers using CA-Buddy for cloud architecture design tasks, to investigate how the system-driven flow influences their experience, design process, and perceived support.

\subsection{Study Design}

We conducted a user study with 60 novice engineers using CA-Buddy. Participants were newly hired engineers at a Japanese company, with diverse technical backgrounds such as backend, machine learning/data science, web frontend, game/XR, mobile, and infrastructure. Most had limited or no prior experience with cloud technologies. Each participant was assigned one of three scenarios—web application, mobile application, or data analytics/machine learning—and asked to design an architecture using CA-Buddy. The study included a 20-minute system orientation, a 30-minute design task, and a 10-minute post-task questionnaire. We collected free-text responses and conducted thematic analysis~\cite{braun2006using} to identify key patterns in user perceptions. Two authors independently coded the responses using an inductive approach, developing themes directly from participant input without a predefined coding scheme. Through discussion and iterative refinement, we organized the themes into categories reflecting cognitive and architectural support, learning use cases, and areas for improvement.

We used Gemini 2.0 Flash as the language model for CA-Buddy. Based on prior research showing both limitations in external knowledge integration and advantages of hybrid systems, we enhanced the system by (1) enabling Gemini’s grounding function to retrieve real-time information, and (2) adding a built-in chat interface for natural language discussion during design.
\subsection{Feedback Analysis}
\begin{figure}
    \centering
    \includegraphics[width=1.0\linewidth]{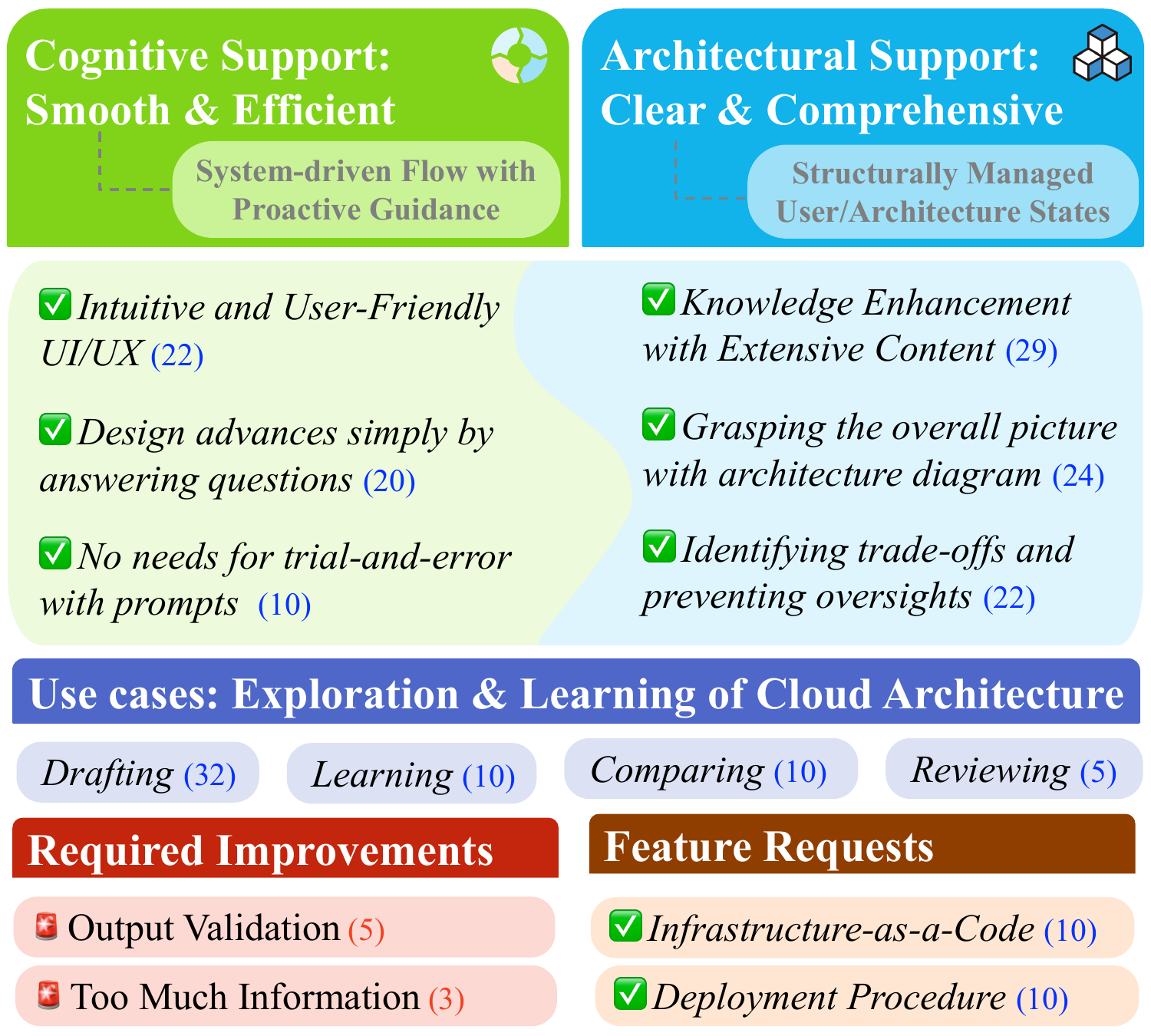}
    \caption{\small Thematic summary of feedback from 60 novice users. Numbers in parentheses indicate the number of mentions.}
    \label{fig:feedback-overview}
\vspace{-1em} 
\end{figure}

Figure~\ref{fig:feedback-overview} summarizes the main themes from user feedback.
Feedback can be grouped into two categories: \textbf{Cognitive Support} enabling smooth and efficient design and \textbf{Architectural Support} ensuring clear and comprehensive design. Users reported benefits such as user-friendly UI/UX (22 mentions), smooth design progress through answering system-generated questions (20), and no need to write detailed prompts (10).
The tool further aided comprehensive architectural work by presenting structured information, supplementing understanding with extensive content (29), offering visual architecture diagrams (24), and helping examine trade-offs and avoid oversight (22).
The chat feature was also valued for asking about unfamiliar services and discussing the proposed architecture (6).

The most common use case was initial architectural drafting (32). Users also reported using the tool to compare different architectures when assessing advantages and disadvantages (10), and to review existing designs for possible oversights or alternative options (5). Notably, several users identified the tool’s educational potential for deepening their understanding of cloud services (10).

Some feedback highlighted areas for improvement, especially the need to validate the generated contents (5) and the difficulty of processing large volumes of information (3).
Users also requested practical features such as Infrastructure-as-Code (IaC) generation and explanations of deployment procedures (10).

\begin{table}[t]
\centering
\small
\caption{\small User feedback: strengths and improvement requests}
\label{tab:feedback-examples}
\begin{tabular}{lp{0.85\linewidth}}
\toprule
\multicolumn{2}{l}{\textbf{Knowledge Support for Novices}} \\
\texttt{FB1} & \textit{The tool automatically created a basic design, which made the whole process much easier. As a beginner, I felt more confident moving forward with the design.} \\

\texttt{FB2} &  \textit{Since I have little knowledge of infrastructure, I wouldn't even know where to start with designing a cloud architecture. Just having the system create an initial draft for me was very helpful.} \\ \midrule

\multicolumn{2}{l}{\textbf{Process Support for Novices}} \\
\texttt{FB3} & \textit{Since it provides an overview, the relevant architectural components, and potential concerns, I hardly needed to write prompts, which made it extremely convenient.} \\

\texttt{FB4} & \textit{It's really helpful that just by inputting my requirements, I can get all the information I need, without having to carefully craft prompts, unlike with ChatGPT.} \\

\texttt{FB5} & \textit{With regular chat-based generative AI, I spend a lot of time crafting prompts to get results like those from CA-Buddy, and even then, things don’t always go smoothly.} \\

\texttt{FB6} & \textit{I really liked that the AI asked follow-up questions to prevent me from making designs based on incorrect assumptions.} \\  \midrule

\multicolumn{2}{l}{\textbf{Learning through Iterative Exploration}} \\
\texttt{FB7} & \textit{Choosing services and making improvements feels like a game or simulator, which helps deepen your understanding as you go.} \\

\texttt{FB8} & \textit{Helpful for learning to create and compare multiple architectures.} \\

\texttt{FB9} & \textit{Even without actual deployment, being able to try out and explore multiple design patterns is a great way to learn.} \\

\texttt{FB10} & \textit{I typically went with ECS, but found that Lambda and API Gateway can handle small or short tasks just fine.} \\ \midrule

\multicolumn{2}{l}{\textbf{Required Improvements and Feature Requests}} \\
\texttt{FB11} & \textit{Since I'm new to infrastructure, I felt a bit overwhelmed by all the information. Concise summaries for each service would be very helpful, with more detailed information available as an option.} \\

\texttt{FB12} &  \textit{Providing links to detailed documentation for each tool (e.g., Amazon S3) would be helpful, and using official icons in the architecture diagrams would make them easier to grasp.} \\

\texttt{FB13} & \textit{As someone unfamiliar with infrastructure, I was not always confident about whether the final design was truly correct.} \\

\texttt{FB14} & \textit{Features such as Terraform code export, cost estimation, and deployment steps would make the tool more practical.} \\

\texttt{FB15} & \textit{It would be even more helpful if there were a table for easy side-by-side comparison of multiple services.} \\
\bottomrule
\end{tabular}
\vspace{-1em} 
\end{table}

Table~\ref{tab:feedback-examples} presents representative user feedback selected to illustrate the main topics identified in Figure~\ref{fig:feedback-overview}.
Participants reported that the system-driven workflow helped fill gaps in their cloud architecture knowledge, which in turn allowed them to engage in the design process with less uncertainty and greater ease (\texttt{FB1}).
In particular, many participants felt that having the system generate an initial architecture proposal could be helpful for overcoming the common hurdle of \emph{not knowing where to start} (FB2).

Participants frequently reported that CA-Buddy’s design process enabled them to smoothly progress through all stages of architectural design, from specifying initial requirements to identifying potential concerns, since the system autonomously provided architectural elements, summaries, and potential issues without requiring users to craft prompts or manage the interaction themselves (\texttt{FB3}). This aspect was particularly appreciated by participants who found prompt creation with conventional chat-based tools to be difficult and time-consuming (\texttt{FB4}, \texttt{FB5}).
Participants also highlighted that the system’s ability to proactively ask targeted follow-up questions was effective in uncovering missing information and preventing misunderstandings, thereby reducing cognitive load and supporting those less familiar with architectural design practices (\texttt{FB6}).

Participants also recognized the educational value of CA-Buddy’s interactive, simulation-like design process. Some described being able to experiment with architectural options much like playing a game or using a simulator (\texttt{FB7}), which made it easy to compare different patterns (\texttt{FB8}), iterate repeatedly to deepen their understanding (\texttt{FB9}), and discover unfamiliar services or possibilities in the process (\texttt{FB10}).
These findings suggest that system-driven approach can not only assist novices with limited experience in designing cloud architectures, but also foster their development of practical design skills through exploration and feedback~\cite{yeh2025bridging}.

Participants reported several issues. Many felt overwhelmed by the volume of information presented, particularly the lengthy textual descriptions. Novice participants found it difficult to identify essential information, requesting more concise summaries, clearer diagrams, and links to additional documentation (\texttt{FB11}, \texttt{FB12}). Some also expressed uncertainty about the validity or suitability of the generated architectures, and asked for better support in evaluating system suggestions (\texttt{FB13}).
This was especially challenging for novices, as limited background knowledge made it difficult to judge the correctness of system outputs. 

As feature requests, user mentioned a desire to move beyond conceptual design to actual implementation.
Specifically, automatic generation of Infrastructure-as-Code (IaC), cost estimation, and deployment instructions are demanded (\texttt{FB14}), and also a comparison table of multiple architectures would help users to comprehend multiple options over given requirements (\texttt{FB15}). Although CA-Buddy supported conceptual learning and design, users indicated that stronger integration with implementation workflows would help bridge the gap from idea to practice.

\section{Related Work}
Recent research demonstrates that LLMs support many aspects of software engineering, from code generation to requirements elicitation~\cite{ramakrishna-2024-codeplan, White2024}. LLM-based systems have also enabled guided interactions for non-experts in educational and analytics domains~\cite{zheng2022exekg, tu2023littlemu}. For cloud architecture design, ArchMind is a chat-based tool that incorporates architectural pattern knowledge and tracks design decisions, helping novices organize and explain the reasoning behind their choices~\cite{diazpace2024helping}.
CA-Buddy instead adopts a structured workflow approach, distinguishing itself from prior chat-based methods.

LLMs increasingly support novices throughout various stages of software development. They can improve task completion and confidence in introductory programming, but sometimes at the cost of deeper understanding~\cite{kazemitabaar2023ai, prather2024widening, hellas2023responses}. LLM-based tools can provide guidance, hints, or explanations not only for basic tasks but also for higher-level activities such as debugging, requirements clarification, and program design~\cite{nguyen2024misread, zamfirescu2025beyond, codeaid2024}. However, these tools can introduce errors, promote over-reliance, or cause information overload if not properly scaffolded~\cite{prather2024widening}. Their effectiveness for novices depends on structured workflows and interaction design~\cite{Sangho2024}.

Recent research has explored how LLM-based tools can support learning in software engineering. For example, integrating conversational assistants into development environments has been shown to improve code comprehension by providing timely explanations~\cite{nam2024using}. Dialog-based scaffolding can also help users acquire more effective prompting skills~\cite{yeh2025bridging}. However, prior studies report challenges such as automation bias and shallow understanding in generative model outputs~\cite{prather2023weird,prather2024widening,zi2025understand}.
Bringing together LLMs, software engineering, and education is an emerging area with significant promise. Our findings suggest system-driven cloud design holds strong potential as novice educational support.

\section{Discussion and Conclusion}
This study found that a system-driven, workflow-based tool can effectively support novice engineers in cloud architecture design. By surfacing missing requirements, relevant trade-offs, and guiding decisions step by step, the tool addressed challenges stemming from limited domain knowledge and difficulties interacting with large language models. In addition to these cognitive and architectural supports, many participants noted that iterative use of the tool promoted incidental learning, such as discovering unfamiliar services and deepening understanding of cloud architecture. However, the study has limitations. All participants were novice engineers from a single company, which may limit the broader relevance of the findings. Moreover, the 30-minute task may not reflect the complexity of real-world design. Participants identified needs for more adaptive information delivery, closer integration with implementation workflows, and mechanisms for verifying the tool’s suggestions. Some expressed uncertainty about the appropriateness of generated designs, highlighting risks of over-reliance or automation bias. Future work should improve clarity and reliability of system outputs—for example, by presenting rationale, linking to documentation, or incorporating user feedback.

\section*{GenAI Usage Disclosure}
Generative AI tools were employed in two primary capacities during the course of this research. First, the research prototype used in the study incorporates Google’s Gemini 2.0 Flash as its backend large language model. This model was selected due to its low-latency response times, which were critical for enabling participants to engage in iterative exploration within the limited time constraints of the user study. Second, OpenAI’s ChatGPT was used during the manuscript preparation phase to improve grammar and enhance linguistic clarity. All authors have verified and are fully accountable for the content of the paper.

\balance
\bibliographystyle{ACM-Reference-Format}
\bibliography{references}

\end{document}